\newif\iflayout
\layouttrue
\iflayout
   \documentstyle[aps,epsfig,multicol]{revtex}
\else
   \documentstyle[preprint,eqsecnum,aps]{revtex}
\draft
\fi

\draft

\begin{document}

\title{
Disorder, Order, and Domain Wall Roughening in the 2d Random Field Ising
Model}

\author{E.~T.~Sepp\"al\"a,$^1$ V.~Pet\"aj\"a,$^1$ and M.~J.~Alava$^{2,1}$}

\address{$^1$Helsinki University of Technology, Laboratory of
Physics, P.O.Box 1100, 02015 HUT, Finland}
\address{$^2$Nordic Institute for Theoretical Physics, Blegdamsvej 17, 
DK-2100 Copenhagen, Denmark}

\maketitle

\begin{abstract}
Ground states and domain walls are investigated 
with exact combinatorial optimization in two-dimensional 
random field Ising magnets.
The ground states break into domains above a length scale that depends
exponentially on the random field strength squared. 
For weak disorder, this paramagnetic structure has
remnant long-range order of the percolation type.
The domain walls are super-rough in ordered systems 
with a roughness exponent $\zeta$ close to 6/5. 
The interfaces exhibit rare fluctuations and multiscaling
reminiscent of some models of kinetic roughening and hydrodynamic
turbulence.
\end{abstract}

\noindent {\it PACS \# \ \ 64.60.Cn, 05.50.+q, 64.60.-i }
%\newpage
\iflayout
   \begin{multicols}{2}[]
\narrowtext
\fi

The random field Ising model attracts interest since
it presents an example of competing mechanisms for order and
disorder. The local spin couplings favor ferromagnetic ordering
whereas variations in the random fields favor disorder. This
competition affects thermodynamic properties. Disordered systems 
are very often governed by the zero-temperature behavior, or the 
structure and energy of the ground state (GS). In the random 
field Ising model (RFIM) statistical mechanics of interfaces or domain 
walls becomes the key question. This is important for finite-temperature 
dynamics, for coarsening, aging, and fluctuations. The interest is
in the scaling and universality in a problem dominated by a
complicated, multivalley energy landscape \cite{HaH95,Nat97}.

The effect of dimensionality on order or disorder was solved
by Aizenman and Wehr. They proved rigorously that the two-dimensional (2D)
RFIM has in the thermodynamic limit no long-range ferromagnetic 
character \cite{Aiz89}. 
In 3D order was shown to exist at finite temperatures and weak
fields \cite{Bri87}, and therefore the lower critical dimension
of the RFIM is 2. Here we study the transition from order to
disorder in the 2D RFIM: ground states and
scaling properties of single domain walls with varying
system size.
The expectation is that the ground state 
becomes unstable to domain formation at a {\it breakup length scale}
\cite{Bin83} because of domain wall entropy even at zero temperature.
In contrast, a simple energy or Imry-Ma domain argument indicates 
for weak fields in two dimensions long-range order 
(if $hL^{d/2} < JL^{d-1}$, where $h$ is the RF strength, $J$ the
strength of ferromagnetic couplings, $L$ length scale, and $d$ the
dimension) \cite{Imr75,Nat97}.  This fails so 
that domains, albeit large ones, do exist for arbitrarily weak
fields.  Existing finite temperature Monte 
Carlo \cite{Fer85} and exact ground state results
\cite{Ess97} do not extend into this regime.

The scaling properties of domain walls is studied here with the domain wall 
renormalization group (DWRG). One considers DW's imposed
with boundary conditions, and compares to systems without
forced DW's and with the same disorder to find the DW energy.
The domain walls are predicted to be self-affine
by functional renormalization group calculations and an Imry-Ma argument, 
with the roughness exponent 
$\zeta_{d}^{RF}= (5-d)/3$ \cite{Fis86,Moo96}.
$\zeta_d^{RF}$ shows, e.g., how the interface width scales, $w \sim L^\zeta$ 
($w^2 = \langle z^2 - {\bar z}^2\rangle$, where $z$ is the local interface
height).
In this picture $\zeta$ vanishes at the upper 
critical dimension and at $d=2$ $\zeta_2 = 1$ \cite{Bin83,Moo96}.
The domain wall energy would concurrently be expected to be linear,
$E(L) = E_0 L + E_1L^\theta$. The energy fluctuation exponent 
$\theta$ should obey the exponent relation $\theta = 2 \zeta + d -3$, 
similar to the random bond Ising model and directed polymers 
\cite{Hus85,HaH95}.  The 1+1 -dimensional RF domain wall problem maps 
in the continuum limit directly to the Kardar-Parisi-Zhang or Burger's 
equation, the paradigm of interface models in disordered media \cite{Kar86},
so that again $\zeta_{2} = 1$ and $\theta_{2} = 1$ \cite{Zha86}.  

The picture of self-affine DW's has been claimed to be 
confirmed by both early transfer-matrix calculations \cite{Fer83} and 
studies using combinatorial optimization \cite{Moo96,Jos97}.  
In this Rapid Communication this is shown to be false. The
domain walls exhibit rich scaling 
reminiscent of turbulent behavior as in certain kinetic growth models
\cite{Kru94,ESS} and in ordinary hydrodynamics \cite{Kruref,kirja}. 
Also, the concept of self-affinity is not valid because of the 
length scale induced by ground state breakup.

Finding the ground state of the
RFIM maps exactly into the minimum cut -- maximum flow problem of network or
combinatorial optimization \cite{Aur85}. The use of such algorithms, 
pioneered by Ogielski \cite{Ogi86}, has recently started become common 
as one can do exact disorder averages for systems governed by zero-temperature
and energy landscape effects \cite{ala}.  
A related problem solvable with the method is the DAFF (diluted antiferromagnet
in a field) providing an experimental realization.

The application of combinatorial optimization starts by
augmenting the RFIM with two extra sites. The 
network optimization problem is defined on a graph,
in which each edge corresponds to a site in the augmented
RFIM. Each of the original sites is connected with one of the two 
extras, depending on the sign of the local field $h_i$. The
capacities of the vertices in the graph are equal to either $2J$
or $2 |h_i |$ for couplings to the extra sites.  
This is a network flow problem, since the connections equal
local flow constraints or capacities. The maximum flow between the extra 
sites gives the ground state energy, and the division
to two spin states among the Ising spins is the minimum
cut that results in the maximum flow. This method is exact and does not 
suffer from metastability like normal Monte Carlo or optimization
with simulated annealing.  We use an efficient push-relabel preflow-type 
\cite{Goltar88} code.  The CPU-time scales  as $t_{CPU} \sim N^{1.2}$, 
with $N \sim L^2$ increasing thus almost linearly in $N$.  
Systems can be studied up to $L=1000$ ($N = 10^6$).

Figure~\ref{fig1} and the inset show examples of ground states with
weak and strong disorder and with domain wall-enforcing or periodic
boundary conditions, respectively \cite{comm}. The random fields $h_i$
obey either a bimodal distribution ($P(h_i) = \frac{1}{2}
[\delta(h_i-\Delta ) + \delta(h_i+\Delta)]$) or a Gaussian one,
($P(h_i) = 1/(\sqrt{2 \pi} \Delta)
\exp[\frac{1}{2}(h_i/\Delta)^2]$, $\Delta$
measures the standard deviation, $J=1$).  First we characterize the
transition from the case of Fig.~\ref{fig1}, a ferromagnetic
ground state here with an imposed domain wall, to that shown in the inset,
with a negligible magnetization. Then the properties of
single domain walls are studied.

We make the assumption of one single length scale, 
proportional to that at which the order vanishes
(e.g., the magnetization becomes zero).  We measure the probability of
a purely ferromagnetic GS, $P_{FM}(L) = P(L, |m| = 1)$, 
as a function of $L$ with fixed $\Delta$.  
This probability maps for both types of disorder to the
magnetization [$m = m(P_{FM})$]. The break-up length scale is
defined with $P_{FM}(L) = 0.5$. The advantage is that the breakup of the
ground state is visible at much smaller $L$ than with other
order parameters, making it possible to study
breakup to a domain structure with $L \rightarrow
\infty$. Other choices could be the cluster size
distribution, the spin-spin correlation length, 
magnetization, and so on.  For example, the correlation length 
shows finite size effects, which might be partly explained below. 

$L_b$ for this definition is depicted for varying $\Delta$ in
Fig.~\ref{fig2}. The prediction that the 2D RFIM ground state should
have no long-range order is based on the fact that at large enough scales
entropy, the many possible configurations available should make the
domain wall energy vanish \cite{Bin83}. Our results yield,
in agreement, an exponential length scale
\[
L_b \sim \exp{(A[1/\Delta ]^2)},
\]
where the disorder-dependent constant $A = 1.9 \pm 0.2$ and $2.1
\pm 0.2$ for bimodal and Gaussian disorder, respectively. The
definition of $L_b$ implies that the magnetization vanishes at a
larger $L > L_b$.  The values of $A$ are different from the ones obtained
by finite-temperature Monte Carlo simulations for small $L$
\cite{Fer85}.  These results prove that the
mechanism for the breakup of the GS is due to entropic effects.

No ferromagnetic order exists in the ensuing domain structure with 
zero magnetization. For strong disorder one can show that the
spin-spin correlation length is proportional to the average cluster
size for both $L< L_b$ and $L \ge L_b$. Here we study the disorder
averaged properties of the largest clusters.  These are found to
percolate and thus give rise to sub-dominant (the weight of the
spanning cluster vanishes in the thermodynamic limit) {\it long-range
order}.  For bimodal disorder the fractal dimension is $d_f =
1.90 \pm 0.02$ (Fig.~\ref{fig2} inset), very close to the exact value of 
standard 2D percolation 91/48. The inset of Fig.~\ref{fig2} also shows 
the sum over the random fields of the percolation clusters. This sum scales
with the same fractal dimension 1.90.  Thus the Imry-Ma argument is
not true for the largest clusters as the
global optimization produces domains whose magnetization
is extensive. To summarize for weak disorder there is hidden order in
the ground state of the RFIM structure in two dimensions.  This is not in
contradiction to the exact Aizenman-Wehr-theorem since $m \rightarrow 0$.  
However, it gives
rise to nontrivial correlations in the structure, thus order. For
stronger fields there is a crossover to site percolation
and a nonpercolating structure (as $p_c \sim 0.593$ on a square
lattice and now $p=0.5$).  
The critical $\Delta_c$, below which lattice
effects are smeared out, is $(h/J)_c =2$ for bimodal disorder and
$\Delta_c = 2.1 \pm 0.2$ for Gaussian, respectively. The threshold for
the Gaussian case is a rough estimate. It would interesting but hard to
analyze this percolation transition in detail, since one needs $L > L_b$.

Next we turn to interface scaling. Fig.~\ref{fig3} shows the
interface width, the interface energy $E$, and the energy fluctuations
$\Delta E^2 = \langle E^2-\bar{E}^2\rangle$ up to $L= 500 - 1000$.
$E$ is obtained in the DWRG sense by subtracting the energy of a
ground state from one with an imposed domain wall and identical
disorder.  $\langle \dots \rangle$ is the average over disorder.  
We take the Solid-On-Solid (SOS) -limit: in the case of a multiply
valued interface the highest location is chosen from the 
exact interface configurations. The weight of overhangs is
negligible for weak disorder and small systems. 
In the weak-disorder regime the global roughness exponent is
found as $\zeta \simeq 1.2 \pm 0.05$. As $\zeta > 1$ the RF interfaces 
are super-rough. This is, however, true
only up to a length scale, below which the GS has already broken down
(see inset, Fig.~\ref{fig1}). Above that scale a domain wall becomes
fractal, and $\zeta =1$.  There is a sharp transition between these
two regimes, and the data for $\Delta = 10/9$ in the inset of
Fig.~\ref{fig3}(a) has two regimes corresponding to ``ferromagnetic'' and
``disordered'' ground states.  Figure~\ref{fig3}(b) shows the DWRG result
for the DW energy: there is a logarithmic correction to the DW energy
in the FM phase.  In the paramagnetic phase the energy has only a remnant
contribution from the boundary conditions. For the FM phase the energy
fluctuation exponent is $\theta \simeq 1$. The values for the
exponents $\zeta$ and $\theta$ disagree with the exponent
relation $\theta_{2d} = 2 \zeta_{2d} -1$ \cite{Hus85}.

If one studies interfaces based either on a mapping to the
Burgers'-KPZ equation or functional RG calculations these depend
on the small slope approximation, which is a problem if $\zeta \simeq 1$.  
Fig.~\ref{fig4} shows the statistics of
interface fluctuations in the form of the interface step height
probability density function (PDF) $f(\Delta z_{i,i+1},L)$. $\Delta
z_{i,i+1}$ is the height difference between two neighboring
sites ($z_i$) along the SOS interface ($i = 1,\dots,L$).  The $f(\Delta z,
L)$ show stretched exponential behavior. The PDF's are clearly $L$ dependent, 
but only up to the breakup length scale for interfaces. 
The height differences resemble velocity gradients
or energy dissipation in fluid turbulence and behavior 
in interface growth problems \cite{Kru94,ESS,Gal,Kruref} 
that are governed by intermittent, rare events. 
The step height fluctuations are not restricted
to the exact interfaces. A SOS transfer matrix calculation
(allowing for $\Delta z > 1$) reproduces these features, and
demonstrates that the interpretation of Ref. \cite{Fer83} is
wrong since the true scaling behavior is super-rough at, also, 
low temperatures.  A multifractal study of the average step height 
$|\Delta z|$ and the interface height-height correlation functions 
$G_k (r) = \langle|h(l)h(l+r)|^k\rangle$ indicates that the local
interface scaling is multiaffine, e.g., $G_k(r) \sim
A_k r^{\alpha_k k}$.  For instance, $\alpha_k \simeq 0.88, 
\dots, 0.9$ for $k=1$, but for already $k=2$ $\alpha_k 
\simeq 0.66$ with the exponent being a weak function of $L$ 
at fixed $\Delta$. The higher exponents $\alpha_k$ 
decrease with $k$ and increase with $L$ for moderate $L$. 
Thus, there is another analogy between height-height
correlations of the RFIM and velocity-velocity correlations
in turbulence. Both the amplitude  of $G$ ($A_k$) and 
$|\Delta z|$ do not self-average but scale with $L$ and $\Delta$.
It is tempting to draw a parallel with $L_b$
in here and the outer length scale of turbulence.
In both cases the largest length scale is fixed by external
conditions: the strength of randomness or the Reynolds 
number \cite{Bo96}. The correspondence is not one-to-one, however,
since the scaling properties depend on both $L_b$ and $L$ (Fig.~4).

In conclusion, we demonstrate the breakdown of the ground state,
at zero temperature, in the 2D random field Ising model.
There is however hidden, long-range order in the form 
of the spanning cluster that
seems first contradictory to destroyed ferromagnetic order.
This arises by ``entropic optimization'' so that the cluster
magnetization becomes extensive. The annihilation of order
with increasing sample size is reflected in the properties of
domain walls. For small systems and weak fields the domain walls are
super-rough, with a roughness exponent that is well in excess of
analytic estimates.  This can be traced to ``turbulent'' rare
interface fluctuations, but in an {\it equilibrium} system
in contrast to models of kinetic surface roughening or Navier-Stokes
turbulence. In large systems the concept of an individual domain 
wall becomes ill-defined. The domain wall energy has a logarithmic
correction: one should study how far this lack
of self-affinity penetrates the GS energy landscape properties
and, perhaps, dynamical behavior.
It will be interesting to see if the nonstandard interface
scaling properties persist in higher dimensions or in the
presence of an applied field. We believe that this is so
in the latter case, though the ground states are naturally
ferromagnetic. This would have consequences for driven
interfaces below the crossover to annealed disorder 
for a strong enough driving force \cite{Gal,Dro98}.

This work has been supported by the Academy of Finland (MATRA and M. J. A.).

%%%%%%%%%%%%%%%%%%%%%%%%%%%%%

%%%%%%%%%%%%%
%FIGURES
%%%%%%%%%%%%%

\begin{figure}[f]
\centerline{\epsfig{file=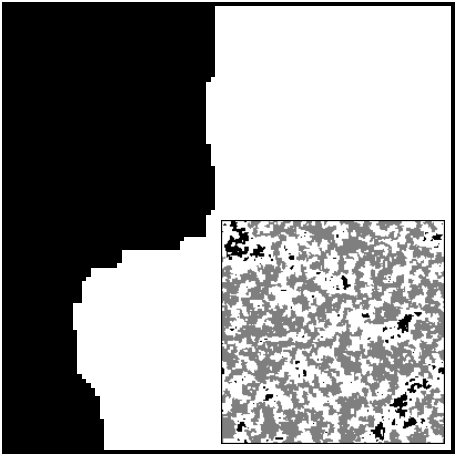,width=6cm}}
\vspace{5mm}
\caption{Two typical examples of RFIM ground states 
with periodic and forced boundary conditions.  Weak bimodal disorder,
$\Delta = 10/17, L=100$. Note the large jumps on the interface and the
lack of overhangs. The interface is the boundary between the
black/white domains (different spin states).  Inset shows a
strong Gaussian disorder case, $\Delta = 2$. Spins that point ``down''
in the ground state are drawn white. The ``up'' spins are black or grey
if in the largest (percolation) cluster.}
\label{fig1}
\end{figure}

\begin{figure}[f]
\centerline{\epsfig{file=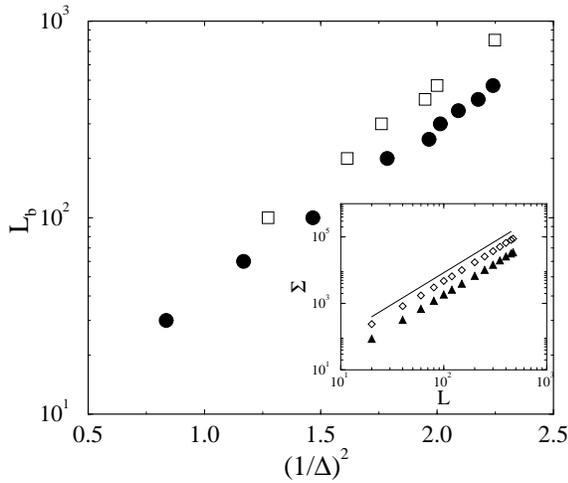,width=65mm,angle=-90}}
\caption{$L_b$ vs. $(1/\Delta)^2$ for bimodal and Gaussian 
disorder (closed circles and open squares, respectively), 
calculated from $P_{FM}(L_b ) = 0.5$. $\Delta = h/J$ for 
binomial and $\Delta = \delta h$ for Gaussian disorder.
The inset shows the average mass of spanning clusters for bimodal 
$\Delta= 25/13$ up to $L=470$ (open diamonds). The plot shows also the 
sum of the random fields of the sites belonging to the same clusters 
(closed triangles). The 2D percolation fractal dimension $d_f =$ 91/48 
is indicated with a line.}
\label{fig2}
\end{figure}

\begin{figure}[f]
\centerline{\epsfig{file=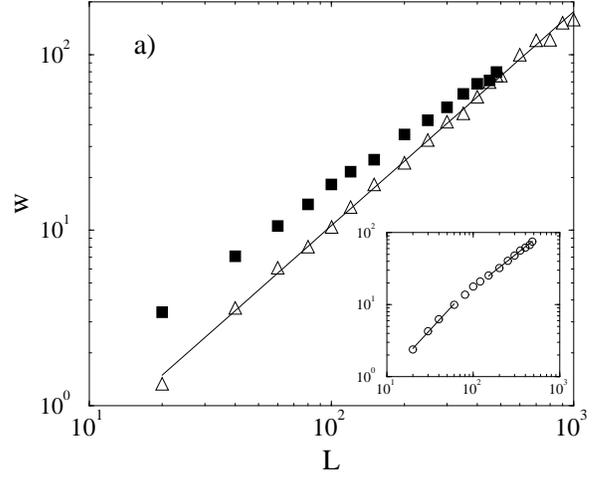,width=65mm,angle=-90}}
\centerline{\epsfig{file=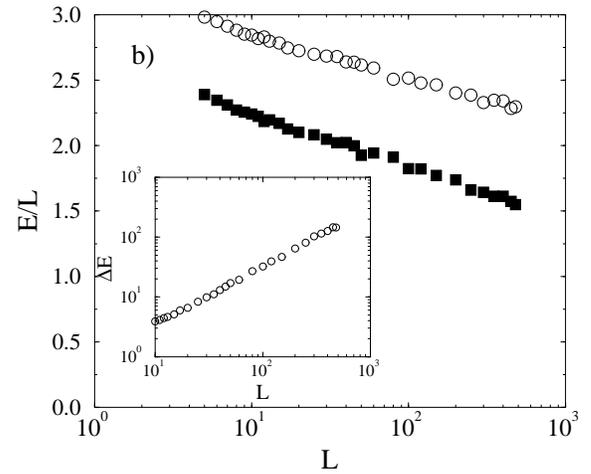,width=65mm,angle=-90}}
\caption{(a) Scaling of the global interface width for 
bimodal disorder. $\Delta =$ 2/3 (open triangles) and 3/2 (closed
squares). The line indicates a least-squares fit with a roughness
exponent $\zeta =$ 1.20 $\pm$ 0.05. The inset shows the crossover in
interface properties with increasing system size ($\Delta =$ 10/9).  
(b) Scaling of the energy (per length) for bimodal $\Delta =$ 1/3 (open
circles) and 5/12 (closed squares). The inset shows the scaling of
energy fluctuations for $\Delta =$ 1/3.}
\label{fig3}
\end{figure}

\begin{figure}[f]
\centerline{\epsfig{file=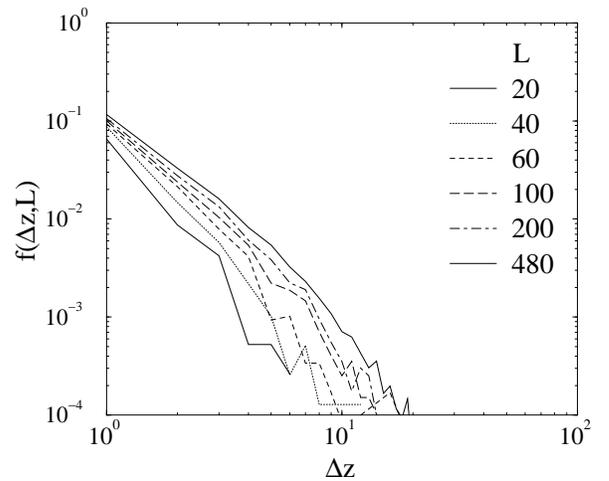,width=65mm,angle=-90}}
\caption{Interface step probability density function for 
$h/J = 1/2$, $L= 20,40,100,200,480$. For simplicity
the data includes only those steps that do not involve
local overhangs.}
\label{fig4}
\end{figure}

\iflayout
  \end{multicols}
\widetext
\else

\fi


\begin{references}
\bibitem{comm} Spins are set up or down at opposing boundaries to
enforce the domain wall, with no regard of the local
fields.
\end{references}
\end{document}